\documentclass[onecolumn,12pt,showpacs,preprintnumbers,amsmath,amssymb,floatfix]{revtex4}

\usepackage{graphicx}
\usepackage{dcolumn}
\usepackage{bm}
\usepackage{amsfonts}

\DeclareMathAlphabet{\mathsfsl}{OT1}{cmr}{bx}{it}
\begin{document}

\title{Rheological study of polymer flow past rough surfaces with slip boundary conditions}

\author{Anoosheh Niavarani and Nikolai V. Priezjev}

\affiliation{Department of Mechanical Engineering, Michigan State
University, East Lansing, Michigan 48824}

\date{\today}

\begin{abstract}

The slip phenomena in thin polymer films confined by either flat or
periodically corrugated surfaces are investigated by molecular
dynamics and continuum simulations. For atomically flat surfaces and
weak wall-fluid interactions, the shear rate dependence of the slip
length has a distinct local minimum which is followed by a rapid
increase at higher shear rates. For corrugated surfaces with
wavelength larger than the radius of gyration of polymer chains, the
effective slip length decays monotonically with increasing
corrugation amplitude. At small amplitudes, this decay is reproduced
accurately by the numerical solution of the Stokes equation with
constant and rate-dependent local slip length. When the corrugation
wavelength is comparable to the radius of gyration, the continuum
predictions overestimate the effective slip length obtained from
molecular dynamics simulations. The analysis of the conformational
properties indicates that polymer chains tend to stretch in the
direction of shear flow above the crests of the wavy surface.

\end{abstract}



\maketitle

\section{Introduction}
\label{sec:Introduction}

The dynamics of fluid flow in confined geometries has gained renewed
interest due to the recent developments in micro- and
nanofluidics~\cite{Darhuber05}. The investigations are motivated by
important industrial applications including lubrication, coating,
and painting processes. The flow behavior at the sub-micron scale
strongly depends on the boundary conditions at the liquid/solid
interface. A number of experimental studies on fluid flow past
nonwetting surfaces have shown that the conditions at the boundary
deviate from the no-slip assumption~\cite{Review05}. The most
popular Navier model relates the slip velocity (the relative
velocity of the fluid with respect to the adjacent solid wall) and
the shear rate with the proportionality coefficient, {\it the slip
length}, which is determined by the linear extrapolation of the
fluid velocity profile to zero. The magnitude of the slip length
depends on several key parameters, such as
wettability~\cite{Churaev84,Cottin02,SchmatkoPRL05,JolyYbertPRL06},
surface
roughness~\cite{HervetLegerPRL00,Granick02,CraigPRL03,Archer03,LegerLang06,Vinograd06},
complex fluid structure~\cite{Migler93,MackayVino}, and shear
rate~\cite{Granick01,CraigPRL01,Breuer03}. However, the experimental
determination of the slip length as a function of these parameters
is hampered by the presence of several factors with competing
effects on the wall slip, e.g. surface roughness and
wettability~\cite{HervetLegerPRL00} or surface roughness and shear
rate~\cite{Granick02}.

In recent years, molecular dynamics (MD) simulations have been
widely used to examine the slip flow past atomically smooth,
homogeneous
surfaces~\cite{KB89,Thompson90,Thompson95,Nature97,Jabbar99,Barrat99,Cieplak01,Phil04,Priezjev04,Priezjev07}.
The advantage of the MD approach is that the velocity profiles and
shear stresses are resolved at the molecular level. The slip length
in the shear flow of simple fluids past crystalline walls is a
function of the wall-fluid density ratio~\cite{Thompson90,Nature97},
the relative size of wall atoms and fluid
molecules~\cite{Nature97,Phil04}, the surface
energy~\cite{Thompson90,Nature97,Priezjev07}, and the interfacial
shear rate~\cite{Nature97,Priezjev04,Priezjev07}. Weak wall-fluid
interactions and incommensurable structures of the solid and fluid
phases at the interface usually lead to enhancement of
slip~\cite{Thompson90,Nature97,Cieplak01,Priezjev07}.  If the slip
length at low shear rates is about several molecular diameters then
it increases with the shear rate, and the slope of the rate
dependence is greater for weaker wall-fluid
interactions~\cite{Nature97,Priezjev07}. The rate dependence of the
slip length in the flow of polymer melts is more complicated because
of the additional length and time scales associated with the
dynamics of polymer chains at the interface and the shear thinning
viscosity~\cite{Thompson95,dePablo96,Koike98,Priezjev04,Priezjev08}.

In the presence of surface
roughness~\cite{Gao2000,Jabbar00,Priezjev06,HartingPRL07,PriezjevJCP07}
or chemical
patterning~\cite{LaugaStone03,HendyPRE05,Priezjev05,ShengPRE05,HendyLund07},
the fluid flow near the solid boundary is perturbed on the length
scales of the surface heterogeneities and its description requires
definitions of the effective slip length and the average location of
the reference plane. Most commonly, the location of the reference
plane is defined as the mean height of the surface asperities, and
the shear rate is determined by averaging of the fluid flow over the
typical length scale of the surface inhomogeneities. In general, the
surface roughness is expected to reduce the effective slip length
for wetting liquids~\cite{Jabbar00,Priezjev06,PriezjevJCP07}. For
sufficiently rough surfaces the no-slip boundary condition can be
achieved even if the local condition is of zero shear
stress~\cite{Richardson73}. However, in special cases, when the
fluid is partially dewetted at the nanostructured, or the so-called
``superhydrophobic'' surfaces, the slip length might be enhanced up
to a few
microns~\cite{BBNatureMat03,CottinEPJE04,SbragagliaPRL06,Sbragaglia07note,Rothstein04,ChoiPRL06,YbertPRL06}.

In recent MD studies on shear flow of simple fluids, the behavior of
the effective slip length was investigated in the Couette cell with
either mixed boundary conditions~\cite{Priezjev05} or periodic
surface roughness~\cite{Priezjev06}. In the first study, the lower
stationary wall with mixed boundary conditions was patterned with a
periodic array of stripes representing alternating regions of finite
slip and zero shear stress. In the other study~\cite{Priezjev06},
the periodically roughened surface was modeled by introducing a
sinusoidal offset to the position of the wall atoms. At the wavy
wall, the local slip length is modified by the presence of curvature
and becomes position-dependent along the curved
boundary~\cite{Einzel90,Einzel92}. A detailed comparison between
continuum analysis and MD simulations shows an excellent agreement
between the velocity profiles and effective slip lengths when the
characteristic length scale of substrate inhomogeneities is larger
than approximately thirty molecular
diameters~\cite{Priezjev05,Priezjev06}. In the case of rough
surfaces, an additional correction due to variable wall density was
incorporated in the analysis~\cite{Priezjev06}. The problem of
applicability of the results obtained for monoatomic fluids to
polymer melts is important for modeling polymer flows in confined
geometries and design of the hybrid continuum-atomistic algorithms.

In this paper, the MD simulations are carried out to study the
dynamic behavior of the slip length at the interface between a
polymer melt and atomically flat or periodically corrugated
surfaces. The MD results for flat crystalline walls confirm previous
findings~\cite{Priezjev08} that the slip length goes through a local
minimum at low shear rates and then increases rapidly at higher
shear rates. For periodically corrugated surfaces and wetting
conditions, the effective slip length decreases gradually with
increasing values of the wavenumber. The solution of the Stokes
equation with either constant or rate-dependent local slip length is
compared with the MD simulations for corrugated surfaces with
wavelengths ranging from molecular dimensions to values much larger
than the radius of gyration of polymer chains. The orientation and
the dynamics of linear polymer chains are significantly affected by
surface roughness when the corrugation wavelengths are comparable
with the radius of gyration.

The rest of the paper is organized as follows. The details of
molecular dynamics and continuum simulations are described in the
next section. The dynamic response of the slip length and shear
viscosity in the cell with atomically flat surfaces is presented in
Sec.\,\ref{MD_large_flat}. The results of MD simulations for
corrugated walls with large wavelengths and comparison with
continuum predictions are reported in
Sec.\,\ref{MD_large_corrugated}. The slip behavior for small
wavelengths and the conformational properties of the polymer chains
near the rough surfaces are analyzed in
Sec.\,\ref{MD_small_corrugated}. The summary is given in the last
section.

\section{The details of the numerical simulations}
\label{sec:details}

\subsection{Molecular dynamics model}
\label{sec:MD_detail}

\begin{figure}[t]
\vspace*{-5mm}
\includegraphics[width=10.4cm,height=9.244444cm,angle=0]{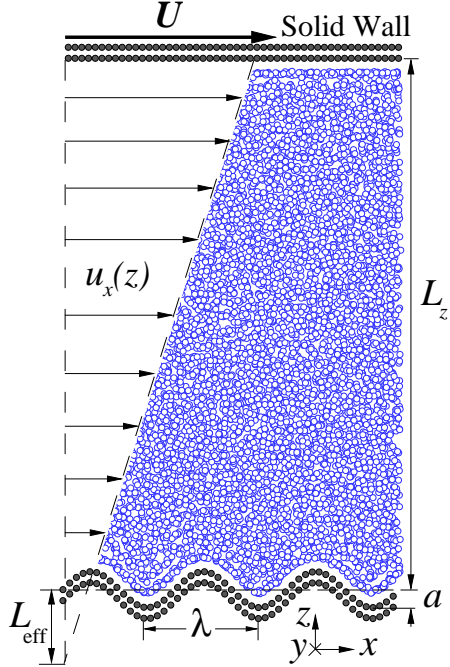}
\vspace*{-1mm}%
\caption{(Color online) A snapshot of the fluid monomers (open
circles) confined between solid walls (closed circles) obtained from
the MD simulations. The atoms of the stationary lower wall are
distributed along the sinusoidal curve with the wavelength $\lambda$
and amplitude $a$. The flat upper wall is moving with a constant
velocity $U$ in the $\hat{x}$ direction. The effective slip length
$L_{\text{eff}}$ is determined by the linear extrapolation of the
velocity profile to $u_{x}\!=0$.} \label{schematic}
\end{figure}

The computational domain consists of a polymeric fluid confined
between two atomistic walls. Figure\,\ref{schematic} shows the MD
simulation setup. Any two fluid monomers within a cut-off distance
of $r_{c}=2.5\,\sigma$ interact through the truncated Lennard-Jones
(LJ) potential
\begin{equation}
V_{LJ}(r)=4\varepsilon\Big{[}\Big{(}\frac{\sigma}{r}\Big{)}^{12}-\Big{(}\frac{\sigma}{r}\Big{)}^{6}\Big{]},
 \label{LJ_potential}
\end{equation}
where $\varepsilon$ is the energy scale and $\sigma$ is the length
scale of the fluid phase. The LJ potential was also employed for the
wall-fluid interaction with $\varepsilon_{wf}=\varepsilon$ and
$\sigma_{wf}=\sigma$. The wall atoms do not interact with each
other, and the wall-fluid parameters are fixed throughout the study.
In addition to the LJ potential, the neighboring monomers in a
polymer chain ($N\,{=}\,20$ beads) interact with the finite
extensible nonlinear elastic (FENE) potential
\begin{equation}
V_{FENE}(r)=-\frac{1}{2}kr_{\!o}^{2}\ln\Big{[}1-\Big{(}\frac{r}{r_{\!o}}\Big{)}^{2}\Big{]},
\label{FENE_potential}
\end{equation}
where $r_{\!o}\,{=}\,1.5\,\sigma$ and
$k\,{=}\,30\,\varepsilon\sigma^{-2}$~\cite{Kremer90}. The MD
simulations were performed at a constant density ensemble with
$\rho\,{=}\,0.88\,\sigma^{-3}$. The total number of fluid monomers
is $N_{f}\!=67200$.

The motion of the fluid monomers was weakly coupled to an external
thermal reservoir~\cite{Grest86}. To avoid a bias in the flow
direction, the random force and the friction term were added to the
equation of motion in the $\hat{y}$ direction~\cite{Thompson90}
\begin{equation}
m\ddot{y_{i}}=-\sum_{i\neq j}\frac{\partial
(V_{LJ}+V_{FENE})}{\partial y_{i}}-m\Gamma \dot{y_{i}}+f_{i}(t),
\end{equation}
where $\Gamma\,{=}\,1.0\,\tau^{-1}$ is a friction constant that
regulates the rate of heat flux between the fluid and the heat bath,
and $f_{i}(t)$ is the random force with zero mean and variance
$2m\Gamma k_{B}T\delta(t)$ determined from the
fluctuation-dissipation theorem~\cite{BoonYip}. Temperature of the
Langevin thermostat is set to $T\,{=}\,1.1\,\varepsilon/k_B$, where
$k_{B}$ is the Boltzmann constant. The equations of motion were
integrated using the Verlet algorithm~\cite{Allen87} with a time
step $\Delta t\,{=}\,0.005\,\tau$, where
$\tau\,{=}\,\sqrt{m\sigma^{2}/\varepsilon}$ is the characteristic LJ
time.

The dimensions of the system in the $xy$ plane, unless specified
otherwise, were set to $L_{x}\,{=}\,66.60\,\sigma$ and
$L_{y}\,{=}\,15.59\,\sigma$. The upper wall was composed of two
layers of an fcc lattice with density
$\rho_{w}\,{=}\,1.94\,\sigma^{-3}$, which corresponds to the
nearest-neighbor distance of $d\,{=}\,0.9\,\sigma$ between wall
atoms in the $(111)$ plane. The lower wall was constructed of two
fcc layers of atoms distributed along the sinusoidal curve with the
wavelength $\lambda$ and amplitude $a$. For the largest wavelength
$\lambda\,{=}\,66.60\,\sigma$, the density of the lower wall
$\rho_{w}\,{=}\,1.94\,\sigma^{-3}$ was kept uniform along the
sinusoid (by including additional rows of atoms parallel to the
$\hat{y}$ axis) to avoid additional analysis of the effective slip
length due to variable wall density~\cite{Priezjev06}. In the
present study, the corrugation amplitude was varied in the range
$0\leqslant a/\sigma\leqslant 12.04$. In the absence of the imposed
corrugation ($a\,{=}\,0$) the distance between the inner fcc planes
is set to $L_{z}\,{=}\,74.15\,\sigma$ in the $\hat{z}$ direction.
Periodic boundary conditions were imposed in the $\hat{x}$ and
$\hat{y}$ directions.


The initial velocities of the fluid monomers were chosen from the
Maxwell-Boltzmann probability distribution at the temperature
$T\,{=}\,1.1\,\varepsilon/k_B$. After an equilibration period of
about $3\times10^{4}\,\tau$ with stationary walls, the velocity of
the upper wall was gradually increased in the $\hat{x}$ direction
from zero to its final value during the next $2\times10^{3}\,\tau$.
Then the system was equilibrated for an additional period of
$6\times10^{3}\,\tau$ to reach steady-state. Averaging time varied
from $10^{5}\,\tau$ to $2\times10^{5}\,\tau$ for large and small
velocities of the upper wall respectively. The velocity profiles
were averaged within horizontal slices of $L_{x}\times
L_{y}\times\Delta z$, where $\Delta z=0.2\,\sigma$. Fluid density
profiles near the walls were computed within slices with thickness
$\Delta z=0.01\,\sigma$~\cite{Priezjev07}.


\subsection{Continuum method}
\label{continuum_detail}

A solver based on the finite element method was developed for the
two-dimensional steady-state and incompressible Navier-Stokes (NS)
equation. The NS equation with these assumptions is reduced to
\begin{equation}
\rho\,(\textbf{u}\cdot\nabla\textbf{u})=-\nabla
p+\mu\nabla^{2}\textbf{u}, \label{N-S}
\end{equation}
where \textbf{u} is the velocity vector, $\rho$ is the fluid
density, and $p$ and $\mu$ are the pressure field and viscosity of
the fluid respectively.

The incompressibility condition is satisfied by a divergence-free
velocity field $\textbf{u}$. In order to avoid the decoupling of the
velocity and the pressure fields in the numerical simulation of the
incompressible flow, the penalty formulation is
adopted~\cite{Pepper}. This method replaces the continuity equation,
$\nabla\cdot\textbf{u}=0$, with a perturbed equation
\begin{equation}
\nabla\cdot\textbf{u}=-\frac{p}{\Lambda}, \label{penalty_formula}
\end{equation}
where $\Lambda$ is the penalty parameter. For most practical
applications, where computation is performed with double-precision
$64$ bit words, a penalty parameter between $10^{7}$ and $10^{9}$ is
sufficient to conserve the accuracy~\cite{Pepper}. In our
simulations the incompressibility constraint was set to
$\Lambda=10^{7}$. The pressure term in Eq.\,(\ref{penalty_formula})
is then substituted into the NS equation. The equation
Eq.\,(\ref{N-S}) can be rewritten as follows:
\begin{equation}
\rho\,(\textbf{u}\cdot\nabla\textbf{u})=\Lambda\nabla(\nabla\cdot\textbf{u})+\mu\nabla^{2}\textbf{u},
\label{penalty_final}
\end{equation}
where the continuity equation is no longer necessary~\cite{Pepper}.
The NS equation is integrated with the Galerkin method using
bilinear rectangular isoparametric elements~\cite{Pepper}.

Four boundary conditions must be specified for the continuum
simulation. Periodic boundary conditions are used for the inlet and
outlet along the $\hat{x}$ direction. A slip boundary condition is
applied at the upper and lower walls. In the local coordinate system
(spanned by the tangential vector $\vec{t}$ and the normal vector
$\vec{n}$), the fluid velocity along the curved boundary is
calculated as
\begin{equation}
u_{t}=L_{0}[(\vec{n}\cdot\nabla)u_{t}+u_{t}/R],
\label{slip_boundary}
\end{equation}
where $u_{t}$ is the tangential component of
$\textbf{u}=u_{t}\vec{t}+u_{n}\vec{n}$, $L_{0}$ is the slip length
at the flat liquid/solid interface, the term in the brackets is the
local shear rate, and $R$ is the local radius of
curvature~\cite{Einzel92}. The radius of curvature is positive for
the concave and negative for the convex regions. For a flat surface,
$R\rightarrow\infty$, the boundary condition given by
Eq.\,(\ref{slip_boundary}) simply becomes the Navier slip law.

The simulation is started by applying the no-slip boundary condition
as the initial guess. Once the equations of motion are solved
implicitly, the local slip velocities at the lower and upper
boundaries are updated using Eq.\,(\ref{slip_boundary}). In the next
step, the equations of motion are solved with the updated slip
velocities used as a new boundary condition. The iterative procedure
is repeated until the solution converges to a desired accuracy. The
convergence rate of the iterative solution remains under control
with the under-relaxation value about $0.001$ for the boundary
nodes. In all continuum simulations, the grids at the lower boundary
have an aspect ratio of about one. The computational cost is reduced
by increasing the aspect ratio of the grids in the bulk region.

The normalized average error value in the simulation is defined as
\begin{equation}
\text{error}=\Big[\sum^{N_{p}}_{i=1}\frac{\mid
\textbf{u}_{i}^n-\textbf{u}_{i}^{n+1}\mid}{\mid
\textbf{u}_{i}^{n+1}\mid}\Big]/N_p, \label{err}
\end{equation}
where $N_p$ is the number of nodes in the system, $\textbf{u}_{i}^n$
is the velocity at the node $i$ and time step $n$, and
$\textbf{u}_{i}^{n+1}$ is the velocity in the next time step. The
typical error in the converged solution is less than $10^{-9}$. At
the boundaries the solution satisfies $u_{t}=L_{local}\frac{\partial
u_{t}}{\partial n}$, where the local slip length is
$L_{local}=(L_{0}^{-1}-R^{-1})^{-1}$~\cite{Einzel92}.


\begin{figure}[t]
\vspace*{-10mm}
\includegraphics[width=10.4cm,height=7.944444cm,angle=0]{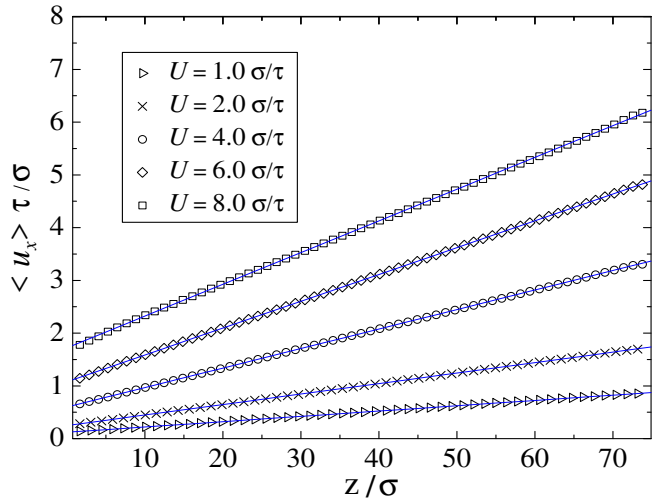}
\vspace*{-5mm}%
\caption{(Color online) Averaged velocity profiles in the cell with
flat upper and lower walls. The solid lines are the best linear fit
to the data. The vertical axes indicate the location of the fcc
lattice planes. The velocities of the upper wall are tabulated in
the inset.} \label{velocity_profiles}
\end{figure}

\section{MD results for flat walls}
\label{MD_large_flat}

The averaged velocity profiles for selected values of the upper wall
speed $U$ are presented in Fig.\,\ref{velocity_profiles}. The
profiles are linear throughout the cell, except for
$U\geqslant6.0\,\sigma/\tau$ where a slight curvature appears in the
region of about $4\,\sigma$ near the walls. Note that the relative
slip velocity at the upper and lower walls increases with increasing
upper wall speed. The shear rate was determined from the linear fit
to the velocity profiles across the whole width of the channel (see
Fig.\,\ref{velocity_profiles}). The uncertainty in the estimated
value of shear rate is due to the thermal fluctuations and the
slight curvature in the velocity profiles near the walls. The
typical error bars for the shear rate are about
$2\times10^{-5}\tau^{-1}$ and $6\times10^{-4}\tau^{-1}$ for small
and large upper wall speeds respectively (not shown).

\begin{figure}[t]
\vspace*{-10mm}
\includegraphics[width=10.4cm,height=7.944444cm,angle=0]{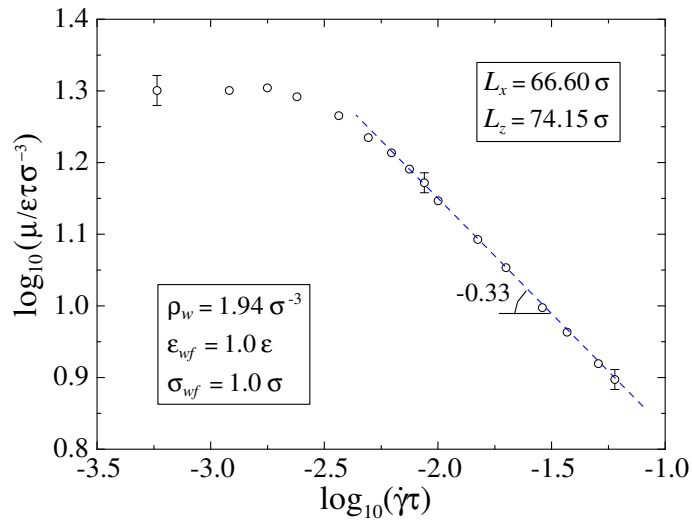}
\vspace*{-5mm}%
\caption{(Color online) Viscosity of the polymer melt
$\mu/\varepsilon\tau\sigma^{-3}$ as a function of shear rate. The
dashed line with the slope $-0.33$ is plotted for reference.}
\label{viscosity_shearrate}
\end{figure}

In this study, the shear stress in steady-state flow was computed
from the Irving-Kirkwood relation~\cite{Kirkwood}. The dynamic
response of the fluid viscosity with increasing shear rate is
presented in Fig.\,\ref{viscosity_shearrate}. At higher shear rates,
the fluid exhibits shear thinning behavior with the slope of about
$-0.33$. Although the power law coefficient is larger than the
reported values in experimental studies~\cite{Bird87}, the results
are consistent with previous MD simulations of polymer melts for
similar flow conditions~\cite{dePablo95,Todd04,Priezjev08}.

The slip length was calculated by the linear extrapolation of the
fluid velocity profile to zero with respect to a reference plane,
which is located $0.5\,\sigma$ away from the fcc lattice
plane~\cite{Priezjev07,Priezjev08}.
Figure\,\ref{sliplength_shearrate} shows the rate dependence of the
slip length in the same range of shear rates as in
Fig.\,\ref{viscosity_shearrate}. The slip length goes through a
shallow minimum at low shear rates and then increases rapidly at
higher rates. The error bars are larger at low shear rates because
the thermal fluid velocity $v^2_{T}\!=k_BT\!/m$ is greater than the
average flow velocity. The nonmonotonic behavior of the slip length
in sheared polymer films with atomically flat surfaces can be
interpreted in terms of the friction coefficient at the liquid/solid
interface, which undergoes a gradual transition from a nearly
constant value to the power law decay as a function of the slip
velocity~\cite{Priezjev08}. The data for the slip length shown in
Fig.\,\ref{sliplength_shearrate} are well fitted by the fourth-order
polynomial
\begin{equation}
L_{0}(x)/\sigma=16.8-72.0\times10\,x+44.0\times10^{3}\,x^2
-97.3\times10^{4}\,x^3+80.5\times10^{5}\,x^4, \label{polynom}
\end{equation}
where $x=\dot{\gamma}\tau$ is the shear rate. The polynomial fit
will be used to specify the boundary conditions for the continuum
solution described in the next section.


\section{Results for periodically corrugated walls: large wavelength}
\label{MD_large_corrugated}

\subsection{MD simulations}
\label{MD_large_vis}

\begin{figure}[t]
\vspace*{-10mm}
\includegraphics[width=10.4cm,height=7.944444cm,angle=0]{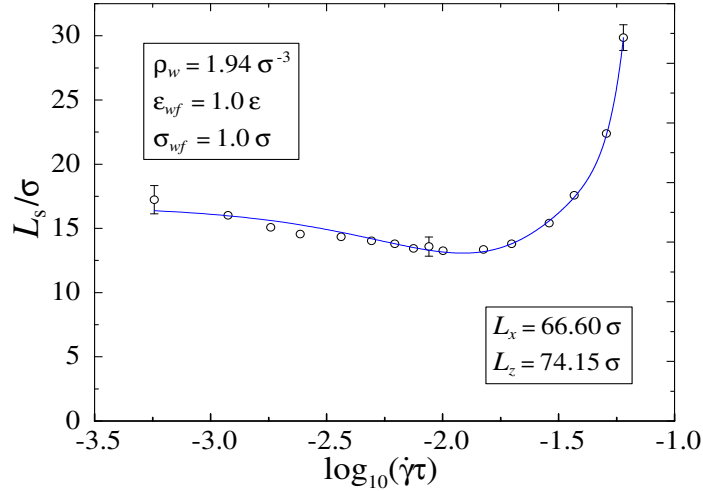}
\vspace*{-5mm}%
\caption{(Color online) Variation of the slip length as a function
of shear rate in the cell with flat upper and lower walls. The solid
line is a fourth-order polynomial fit to the data given by
Eq.\,(\ref{polynom}).} \label{sliplength_shearrate}
\end{figure}

\begin{figure}[t]
\vspace*{-10mm}
\includegraphics[width=10.4cm,height=7.944444cm,angle=0]{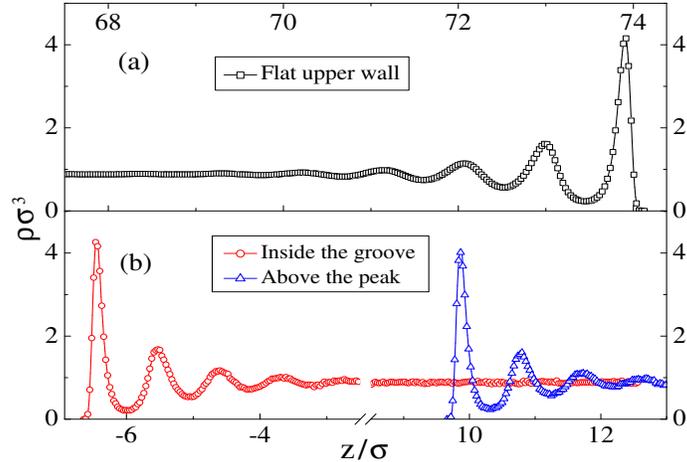}
\vspace*{-5mm}%
\caption{(Color online) Averaged density profiles near the
stationary upper wall (a), above the peak and in the groove of the
lower wall with amplitude $a\,{=}\,8.16\,\sigma$ and wavelength
$\lambda\,{=}\,66.60\,\sigma$ (b).} \label{density_profiles_large}
\end{figure}

The monomer density profiles were computed in the averaging regions
located in the grooves and above the peaks of the corrugated lower
wall with wavelength $\lambda\,{=}\,L_z\,{=}\,66.60\,\sigma$. The
dimensions of the averaging regions are set to $0.5\,\sigma$ and
$L_y\,{=}\,15.59\,\sigma$ in the $\hat{x}$ and $\hat{y}$ directions
respectively. Figure\,\ref{density_profiles_large} shows the monomer
density profiles near the upper and lower walls at equilibrium. The
pronounced density oscillations are attributed to the successive
layering of the fluid monomers near the walls. These oscillations
decay within a few molecular diameters from the walls to a uniform
profile characterized by the bulk density of
$\rho\,{=}\,0.88\,\sigma^{-3}$. Note that the height of the first
peak in the density profile inside the groove is slightly larger
than its value above the crest [see
Fig.\,\ref{density_profiles_large}\,(b)]. The fluid monomers
experience stronger net surface potential in the groove than above
the crest because of the closer spatial arrangement of the wall
atoms around the location of the first density peak in the groove.
This effect is amplified when the local radius of curvature at the
bottom of the grooves is reduced at smaller wavelengths (see below).

\begin{figure}[t]
\vspace*{-10mm}
\includegraphics[width=10.4cm,height=7.944444cm,angle=0]{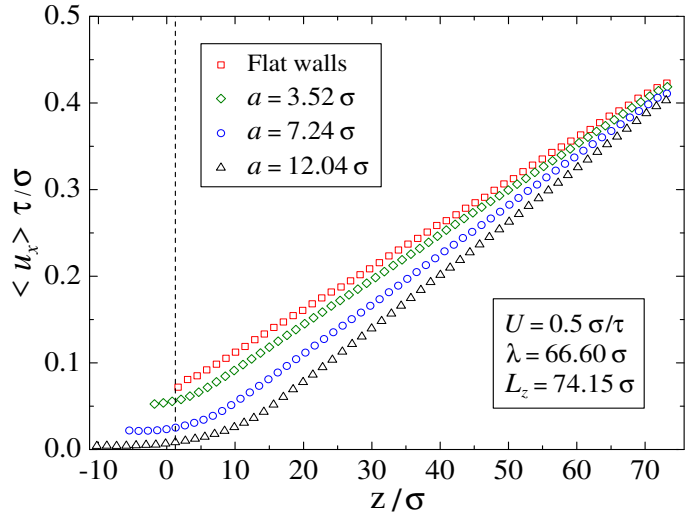}
\vspace*{-5mm}%
\caption{Averaged velocity profiles for the indicated values of the
corrugation amplitude $a$. The vertical dashed line denotes a
reference plane for calculation of the effective slip length at the
corrugated lower wall. The velocity of the flat upper wall is
$U\,{=}\,\,0.5\,\sigma/\tau$.} \label{roughness_MD_large}
\end{figure}

The averaged velocity profiles in the cell with periodically
corrugated lower and flat upper walls are presented in
Fig.\,\ref{roughness_MD_large}. The fluid velocity profiles were
averaged within horizontal slices of thickness $\Delta
z\,{=}\,0.2\,\sigma$ distributed uniformly from the bottom of the
grooves to the upper wall. With increasing corrugation amplitude,
the slip velocity decreases and the profiles acquire a curvature
near the lower wall. Note also that the relative slip velocity at
the upper wall increases with the shear rate. The linear part of the
velocity profiles, $30\leqslant z/\sigma\leqslant60$, was used to
determine the effective slip length $L_{\text{eff}}$ at the
corrugated lower wall. The variation of the effective slip length as
a function of wavenumber $ka\,{=}\,2\,\pi a/\lambda$ is shown in
Fig.\,\ref{MD_continuum_large}. The slip length decreases
monotonically with increasing wavenumber and becomes negative at
$ka\gtrsim0.7$. The results of comparison between the MD and
continuum simulations for three different cases are presented in the
next section.


\subsection{Comparison between MD and continuum simulations}
\label{comp_continuum_large}

In continuum simulations, the length, time and energy scales are
normalized by the LJ parameters $\sigma$, $\tau$ and $\varepsilon$
respectively. The continuum nondimensional parameters are denoted by
the ($\,\tilde{\:}\,$) sign. The size of the two-dimensional domain
is fixed to $66.60\times73.15$ in the $\hat{x}$ and $\hat{z}$
directions, respectively. The following three cases were examined:
the Stokes solution with constant slip length in
Eq.\,(\ref{slip_boundary}), the Stokes solution with
shear-rate-dependent slip length given by Eq.\,(\ref{polynom}), and
the Navier-Stokes solution with constant slip length in
Eq.\,(\ref{slip_boundary}). For all cases considered, the flat upper
wall is translated with a constant velocity $\tilde{U}\,{=}\,\,0.5$
in the $\hat{x}$ direction. Similarly to the MD method, the
effective slip length is defined as a distance from the reference
plane at $a=0$ to the point where the linearly extrapolated velocity
profile vanishes.

\begin{figure}[t]
\vspace*{-10mm}
\includegraphics[width=10.4cm,height=7.944444cm,angle=0]{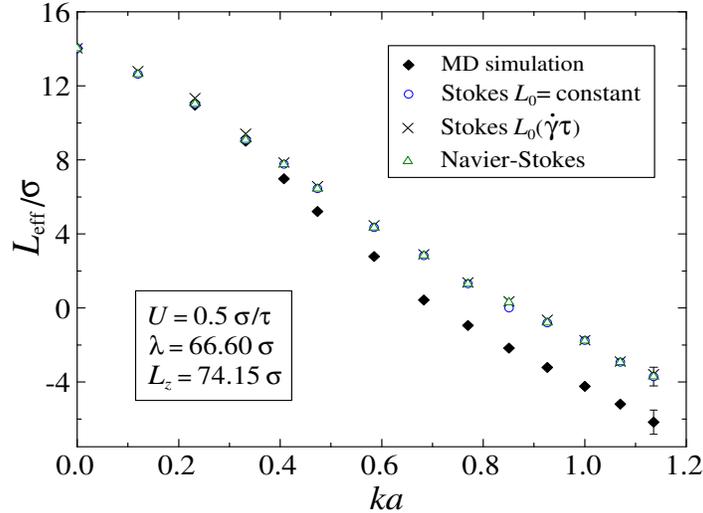}
\vspace*{-5mm} \caption{The effective slip length as a function of
wavenumber $ka$ obtained from the MD simulations ($\blacklozenge$),
the solution of the Stokes equation with rate-independent slip
length $L_{0}$ ($\circ$) and with $L_{0}(\dot{\gamma}\tau)$ given by
Eq.\,(\ref{polynom}) ($\times$), the solution of the Navier-Stokes
equation with $L_{0}$ ($\vartriangle$).} \label{MD_continuum_large}
\end{figure}

In the first case, the finite element method was implemented to
solve the Stokes equation with boundary conditions at the upper and
lower walls specified by Eq.\,(\ref{slip_boundary}) with
$\tilde{L}_0\,{=}\,14.1$. This value corresponds to the slip length
$L_0\,{=}\,14.1\pm0.5\,\sigma$ extracted from the MD simulations in
the cell with flat walls and the velocity of the upper wall
$U\,{=}\,\,0.5\,\sigma/\tau$. Note that at large amplitudes
$\tilde{a}\geqslant8.16$, the normal derivative of the tangential
velocity $\partial u_{t}/\partial n$ at the bottom of the grooves is
negative, while the $\hat{x}$ component of the slip velocity is
positive everywhere along the corrugated lower wall. The dependence
of the effective slip length on the corrugation amplitude is shown
in Fig.\,\ref{MD_continuum_large}. The continuum results agree well
with the approximate analytical solution~\cite{Einzel92} for
$ka\lesssim0.5$ (not shown). For larger amplitudes, $ka>0.5$, where
the analytical solution is not valid, our results were tested to be
grid independent. There is an excellent agreement between slip
lengths obtained from the MD and continuum simulations for
$ka\lesssim0.3$. With further increasing the amplitude, the slip
length obtained from the continuum solution overestimates its MD
value. The results presented in Fig.\,\ref{MD_continuum_large} are
consistent with the analysis performed earlier for simple
fluids~\cite{Priezjev06}, although a better agreement between MD and
continuum solutions was expected at $ka>0.5$ because of the larger
system size considered in the present study.

As discussed in the previous section, the slip length for atomically
flat walls is rate-dependent even at low shear rates (see
Fig.\,\ref{sliplength_shearrate}). In the second case, we include
the effect of shear rate in the analysis of the effective slip
length at the corrugated lower wall and flat upper wall. The Stokes
equation is solved with boundary conditions given by
Eq.\,(\ref{slip_boundary}), where the slip length
Eq.\,(\ref{polynom}) is a function of the local shear rate at the
curved and flat boundaries. The results obtained from the Stokes
solution with constant and rate-dependent slip lengths are almost
indistinguishable (see Fig.\,\ref{MD_continuum_large}). This
behavior can be attributed to a small variation of the intrinsic
slip length Eq.\,(\ref{polynom}) at low shear rates. For example, at
the largest amplitude, $\tilde{a}\,{=}\,12.04$, the local shear rate
at the corrugated wall is position-dependent and bounded by
$|\partial u_{t}/\partial n+u_{t}/R|\leqslant0.0035$. In this range
of shear rates, the normalized value of the slip length in
Eq.\,(\ref{polynom}) varies between
$14.7\leqslant\tilde{L}_0\leqslant16.6$. It is expected, however,
that the effect of shear rate will be noticeable at larger values of
the top wall speed $\tilde{U}$.

In the third case, the Navier-Stokes equation is solved with a
constant slip length $\tilde{L}_0\,{=}\,14.1$ in
Eq.\,(\ref{slip_boundary}) at the flat upper and corrugated lower
walls. The upper estimate of the Reynolds number based on the fluid
density $\tilde{\rho}\,{=}\,0.88$, viscosity
$\tilde{\mu}\,{=}\,20.0$, and the fluid velocity difference across
the channel is $Re\,{\approx}\,1.3$. It was previously shown by Tuck
and Kouzoubov~\cite{Tuck95} that at small $ka$ and
$\tilde{L}_0\,{=}\,0$ the magnitude of the apparent slip velocity at
the mean surface increases due to finite Reynolds number effects for
$Re\gtrsim30$. In our study, the difference between the slip lengths
extracted from the Stokes and Navier-Stokes solutions is within the
error bars (see Fig.\,\ref{MD_continuum_large}). These results
confirm that the slip length is not affected by the inertia term in
the Navier-Stokes equation for $Re\lesssim1.3$. To check how
sensitive the boundary conditions are to higher Reynolds number
flows, we have also repeated the continuum simulations for larger
velocity of the upper wall $\tilde{U}\,{=}\,\,50$, which corresponds
to $Re\,{\approx}\,130$. For the largest corrugation amplitude
$\tilde{a}\,{=}\,12.04$, the backflow appears inside the groove and
the effective slip length becomes smaller than its value for
$\tilde{U}\,{=}\,\,0.5$ by about $0.7$ (not shown).

\begin{figure}[t]
\vspace*{-10mm}%
\includegraphics[width=10.4cm,height=7.944444cm,angle=0]{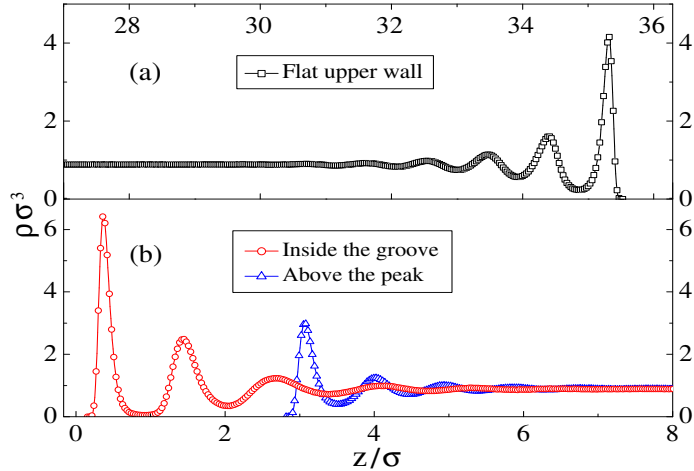}
\vspace*{-5mm}%
\caption{Averaged fluid density profiles near flat upper wall (a)
and corrugated lower wall with amplitude $a\,{=}\,1.4\,\sigma$ and
wavelength $\lambda\,{=}\,7.5\,\sigma$ (b). The velocity of the
upper wall is $U\,{=}\,\,0.5\,\sigma/\tau$.}
\label{density_profiles_small}
\end{figure}

\section{Results for periodically corrugated walls: small wavelengths}
\label{MD_small_corrugated}

\subsection{Comparison between MD and continuum simulations}
\label{comp_continuum_small}

The MD simulations described in this section were performed at
corrugation wavelengths ($\lambda/\sigma\,{=}\,3.75$, $7.5$ and
$22.5$) comparable with the size of a polymer coil. Periodic surface
roughness of the lower wall was created by displacing the fcc wall
atoms by $\Delta z\,{=}\,a \sin(2\,\pi x/\lambda)$ in the $\hat{z}$
direction~\cite{Priezjev06}. In order to reduce the computational
time the system size was restricted to $N_f\!=8580$ fluid monomers
and $L_x\,{=}\,22.5\,\sigma$, $L_y\,{=}\,12.5\,\sigma$ and
$L_z\,{=}\,35.6\,\sigma$. All other system parameters were kept the
same as in the previous section.

\begin{figure}[t]
\vspace*{-10mm}%
\includegraphics[width=10.4cm,height=7.944444cm,angle=0]{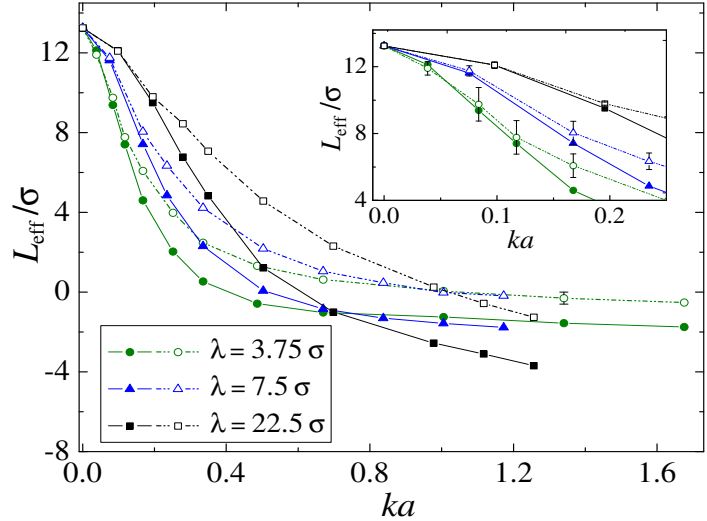}
\vspace*{-5mm}%
\caption{The effective slip length as a function of wavenumber $ka$
for the indicated values of wavelength $\lambda$. Continuum results
are denoted by the dashed lines and open symbols, while the MD
results are shown by straight lines and filled symbols. The inset
shows the same data for $ka\leqslant0.25$.}
\label{MD_continuum_small}
\end{figure}

The representative density profiles near the upper and lower walls
are shown in Fig.\,\ref{density_profiles_small} for the wavelength
$\lambda\,{=}\,7.5\,\sigma$. The height of the first peak in the
density profile is larger in the grooves than near the flat wall or
above the crests of the corrugated surface. The effective slip
length as a function of wavenumber $ka$ is plotted in
Fig.\,\ref{MD_continuum_small}. For all wavelengths, the slip length
decreases monotonically with increasing values of $ka$. At the
smallest wavelength $\lambda\,{=}\,3.75\,\sigma$, the slip length
rapidly decays to zero at $ka\,{\approx}\,0.4$ and weakly depends on
the corrugation amplitude at larger $ka$. Inspection of the local
velocity profiles for $\lambda\,{=}\,3.75\,\sigma$ and
$ka\gtrsim1.0$ indicates that the flow is stagnant inside the
grooves~\cite{Hocking76}.

In the continuum analysis, the Stokes equation with a constant slip
length $\tilde{L}_0\,{=}\,14.1$ in Eq.\,(\ref{slip_boundary}) is
solved for the three wavelengths. The comparison between the MD
results and the solution of the Stokes equation is presented in
Fig.\,\ref{MD_continuum_small}. The error bars are larger for the
smallest wavelength because of the fine grid resolution required
near the lower boundary at $ka\lesssim0.2$ (see inset in
Fig.\,\ref{MD_continuum_small}). The results shown in
Fig.\,\ref{MD_continuum_small} confirm previous findings for simple
fluids~\cite{Priezjev06} that the slip length obtained from the
Stokes flow solution overestimates its MD value and the agreement
between the two solutions becomes worse at smaller wavelengths. It
is interesting to note that the curves for different wavelengths
intersect each other at $ka\,{\approx}\,0.63$ in the MD model and at
$ka\,{\approx}\,1.02$ in the continuum analysis. The same trend was
also observed in the previous study on slip flow of simple fluids
past periodically corrugated surfaces~\cite{Priezjev06}.

\subsection{The polymer chain configuration and dynamics near rough surfaces}
\label{chain_conformation}


In this section, the properties of polymer chains are examined in
the bulk and near the corrugated boundary with wavelengths
$\lambda\,{=}\,3.5\,\sigma$ and $\lambda\,{=}\,7.5\,\sigma$. The
radius of gyration $R_{g}$ was computed as
\begin{equation}
R_{g}^{2}=\frac{1}{N}\sum_{i=1}^{N}(R_{i}-R_{cm})^{2},
\end{equation}
where $R_{i}$ is the three-dimensional position vector of a monomer,
$N\,{=}\,20$ is the number of monomers in the chain, and $R_{cm}$ is
center of mass vector defined as
\begin{equation}
R_{cm}=\frac{1}{N}\sum_{i=1}^{N}R_{i}.
\end{equation}

\begin{table}
\begin{tabular}[t]{|cc|c|c|c|c|c|c|c|c|c|}
\hline
\multicolumn{2}{|l|}{$~~~~~~~\lambda=7.5\,\sigma$}&\multicolumn{3}{l|}{$~~~~~~~~~~~a=0.2\,\sigma$}&\multicolumn{3}{l|}{$~~~~~~~~~~~a=0.6\,\sigma$}&\multicolumn{3}{l|}{$~~~~~~~~~~~a=1.4\,\sigma$}\\
\cline{3-11}
&~~~~&~~~$R_{gx}$~~~&~~~$R_{gy}$~~~&~~~$R_{gz}$~~~&~~~$R_{gx}$~~~&~~~$R_{gy}$~~~&~~~$R_{gz}$~~~&~~~$R_{gx}$~~~&~~~$R_{gy}$~~~&~~~$R_{gz}$~~~\\
\hline\hline
Bulk&Equilibrium&1.18&1.18&1.18&1.18&1.18&1.18&1.18&1.18&1.18\\
&Shear flow&1.70&1.11&1.04&1.76&1.10&1.02&1.79&1.10&1.01\\
\hline
Upper&Equilibrium&1.34&1.37&0.66&1.35&1.36&0.66&1.35&1.35&0.66\\
wall&Shear flow&1.63&1.29&0.63&1.64&1.29&0.63&1.67&1.29&0.63\\
\hline
Peak&Equilibrium&1.35&1.36&0.69&1.40&1.31&0.76&1.46&1.27&0.84\\
&Shear flow&1.65&1.28&0.67&1.85&1.21&0.75&2.17&1.10&0.93\\
\hline
Groove&Equilibrium&1.31&1.37&0.63&1.10&1.47&0.61&0.77&1.69&0.61\\
&Shear flow&1.50&1.34&0.62&1.19&1.44&0.60&0.75&1.89&0.59\\
\hline
\end{tabular}\vspace*{3mm}
\caption{Averaged $\hat{x}$, $\hat{y}$, and $\hat{z}$ components of
the radius of gyration at equilibrium and in the shear flow. The
$R_g$ values are reported in the bulk, near the flat upper wall,
above the peaks, and inside the grooves. The wavelength of the lower
wall is $\lambda\,{=}\,7.5\,\sigma$. The size of the averaging
region inside the grooves and above the peaks is
$\sigma\times12.5\,\sigma\times1.5\,\sigma$. The estimate of the
error bars is $\pm0.03\,\sigma$.} \label{table1}
\end{table}

\begin{figure}[t]
\vspace*{-25mm}%
\includegraphics[width=9.cm,height=8.cm,angle=0]{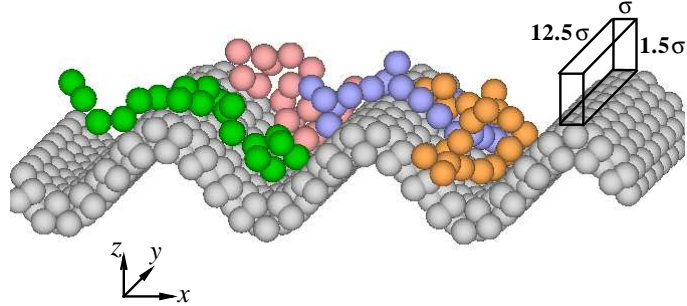}
\vspace*{-20mm}%
\caption{(Color online) A snapshot of four polymer chains near the
lower corrugated wall for wavelength $\lambda\,{=}\,7.5\,\sigma$ and
amplitude $a=1.4\,\sigma$. The velocity of the upper wall is
$U\,{=}\,\,0.5\,\sigma/\tau$.} \label{3Dview_L3.75}
\end{figure}

\noindent The chain statistics were collected in four different
regions at equilibrium ($U\,{=}\,\,0$) and in the shear flow induced
by the upper wall moving with velocity $U\,{=}\,\,0.5\,\sigma/\tau$
in the $\hat{x}$ direction. Averaging regions were located above the
peaks, in the grooves, near the flat upper wall and in the bulk (see
Fig.\,\ref{3Dview_L3.75} for an example). The dimensions of the
averaging regions above the peaks and in the grooves of the lower
wall are $\sigma\times12.5\,\sigma\times1.5\,\sigma$, and near the
upper wall and in the bulk are
$22.5\,\sigma\times12.5\,\sigma\times1.5\,\sigma$. Three components
of the radius of gyration were computed for polymer chains with the
center of mass inside the averaging regions.

In the bulk region, the components of the radius of gyration remain
the same for both wavelengths, indicating that the chain orientation
is isotropic at equilibrium and is not affected by the confining
walls. In the steady-state flow, the effective slip length is
suppressed by the surface roughness and, therefore, the shear rate
in the bulk increases with the corrugation amplitude. This explains
why the $\hat{x}$ component of the radius of gyration $R_{gx}$
increases slightly at larger amplitudes (see
Tables\,\ref{table1}~and~\ref{table2}). Near the upper wall, the
polymer chains become flattened parallel to the surface and slightly
stretched in the presence of shear flow. These results are
consistent with the previous MD simulations of polymer melts
confined between atomically flat
walls~\cite{Bitsanis90,dePablo96,Doi2001}.

\begin{table}
\begin{tabular}[t]{|cc|c|c|c|c|c|c|c|c|c|}
\hline
\multicolumn{2}{|l|}{$~~~~~~~\lambda=3.75\,\sigma$}&\multicolumn{3}{l|}{$~~~~~~~~~~~a=0.07\,\sigma$}&\multicolumn{3}{l|}{$~~~~~~~~~~~a=0.2\,\sigma$}&\multicolumn{3}{l|}{$~~~~~~~~~~~a=1.0\,\sigma$}\\
\cline{3-11}
&~~~~&~~~$R_{gx}$~~~&~~~$R_{gy}$~~~&~~~$R_{gz}$~~~&~~~$R_{gx}$~~~&~~~$R_{gy}$~~~&~~~$R_{gz}$~~~&~~~$R_{gx}$~~~&~~~$R_{gy}$~~~&~~~$R_{gz}$~~~\\
\hline\hline
Bulk&Equilibrium&1.18&1.18&1.18&1.18&1.18&1.18&1.18&1.18&1.18\\
&Shear flow&1.70&1.12&1.03&1.76&1.10&1.02&1.78&1.10&1.01\\
\hline
Upper&Equilibrium&1.33&1.36&0.66&1.36&1.34&0.66&1.37&1.34&0.65\\
wall&Shear flow&1.60&1.32&0.63&1.64&1.29&0.63&1.65&1.27&0.63\\
\hline
Peak&Equilibrium&1.38&1.33&0.67&1.42&1.30&0.71&1.41&1.25&0.92\\
&Shear flow&1.61&1.27&0.65&1.66&1.26&0.69&1.78&1.17&0.84\\
\hline
Groove&Equilibrium&1.33&1.36&0.64&1.25&1.42&0.61&0.48&2.64&0.55\\
&Shear flow&1.60&1.29&0.61&1.60&1.33&0.59&0.47&2.70&0.54\\
\hline
\end{tabular}
\vspace*{1mm} \caption{Averaged $\hat{x}$, $\hat{y}$, and $\hat{z}$
components of the radius of gyration at equilibrium and in the shear
flow. The $R_g$ values are reported in the bulk, near the flat upper
wall, above the peaks, and inside the grooves. The wavelength of the
lower wall is $\lambda\,{=}\,3.75\,\sigma$. The dimensions of the
averaging region are the same as in Table\,\ref{table1}.}
\label{table2}
\end{table}


In the case of a rough surface with the wavelength
$\lambda\,{=}\,7.5\,\sigma$, a polymer chain can be accommodated
inside a groove (see Fig.\,\ref{3Dview_L3.75} for an example). With
increasing corrugation amplitude, the polymer chains inside the
grooves elongate along the $\hat{y}$ direction and contract in the
$\hat{x}$ direction (see Table\,\ref{table1}). The tendency of the
trapped molecules to orient parallel to the grooves was observed
previously in MD simulations of hexadecane~\cite{Jabbar00}. In the
presence of shear flow, polymer chains are highly stretched in the
$\hat{x}$ direction above the crests of the wavy wall. A snapshot of
the unfolded chains during migration between neighboring valleys is
shown in Fig.\,\ref{3Dview_L3.75}. The flow conditions in
Fig.\,\ref{3Dview_L3.75} correspond to a negative effective slip
length $L_{\text{eff}}\approx-2\,\sigma$.

\begin{figure} [t]
\vspace*{-10mm}%
\includegraphics[width=10.4cm,height=9.244444cm,angle=0]{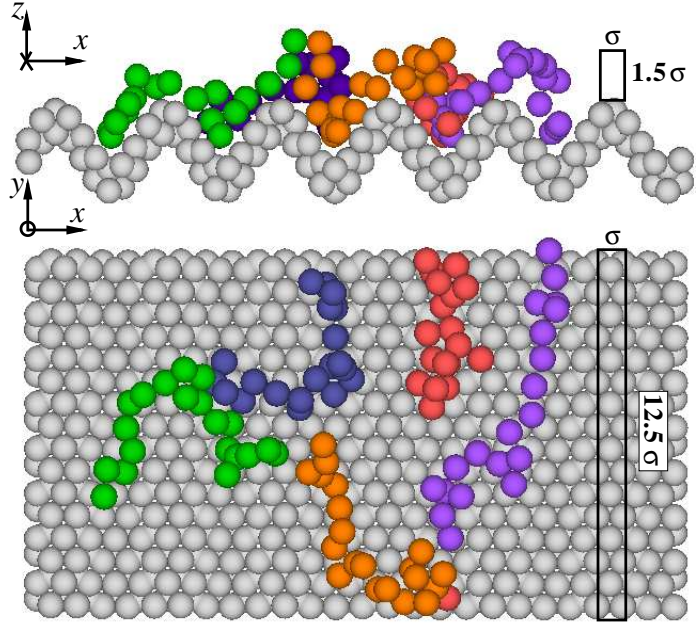}
\caption{(Color online) A snapshot of five polymer chains near the
lower corrugated wall for the wavelength
$\lambda\,{=}\,3.75\,\sigma$ and amplitude $a\,{=}\,1.0\,\sigma$.
The figure shows the side view (top) and the top view (bottom). The
velocity of the upper wall is $U\,{=}\,\,0.5\,\sigma/\tau$.}
\label{3Dview_L1.875}
\end{figure}

For the smallest corrugation wavelength
$\lambda\,{=}\,3.75\,\sigma$, polymer chains cannot easily fit in
the grooves unless highly stretched. Therefore, the $\hat{y}$
component of the radius of gyration is relatively large when the
center of mass is located in the deep grooves (see
Table\,\ref{table2}). Figure\,\ref{3Dview_L1.875} shows a snapshot
of several polymer chains in contact with the lower corrugated wall.
The chain segments are oriented parallel to the grooves and
stretched above the crests of the surface corrugation. Visual
inspection of the consecutive snapshots reveals that the chains near
the corrugated wall move, on average, in the direction of shear
flow; however, their tails can be trapped for a long time because of
the strong net surface potential inside the grooves. For large
wavenumbers $ka\gtrsim0.5$, the magnitude of the negative effective
slip length is approximately equal to the sum of the corrugation
amplitude and $R_{gz}$ of the polymer chains above the crests of the
wavy wall.

\section{Conclusions}

In this paper the effects of the shear rate and surface roughness on
slip flow of a polymer melt was studied using molecular dynamics and
continuum simulations. The linear part of the velocity profiles in
the steady-state flow was used to calculate the effective slip
length and shear rate. For atomically flat walls, the slip length
passes through a shallow minimum at low shear rates and then
increases rapidly at higher shear rates. In the case of periodic
surface heterogeneities with the wavelength larger than the radius
of gyration, the effective slip length decays monotonically with
increasing the corrugation amplitude. For small wavenumbers, the
effective slip length obtained from the solution of the Stokes
equation with constant and shear-dependent local slip length is in a
good agreement with its values computed from the MD simulations, in
accordance with the previous analysis for simple
fluids~\cite{Priezjev06}. At low Reynolds numbers, the inertial
effects on slip boundary conditions are negligible. When the
corrugation wavelengths are comparable to the radius of gyration,
polymer chains stretch in the direction of shear flow above the
crests of the surface corrugation, while the chains located in the
grooves elongate perpendicular to the flow. In this regime the
continuum approach fails to describe accurately the rapid decay of
the effective slip length with increasing wavenumber.

\begin{acknowledgments}

Financial support from the Michigan State University Intramural
Research Grants Program is gratefully acknowledged. The molecular
dynamics simulations were conducted with the LAMMPS numerical
code~\cite{Lammps}. Computational work in support of this research
was performed at Michigan State University's High Performance
Computing Facility.
\end{acknowledgments}

\bibliographystyle{prsty}

\end{document}